\documentclass[10pt,twocolumn,showpacs,amsmath,amssymb,osajnl,floatfix,superscriptaddress]{revtex4-1}

\usepackage{graphicx}
\usepackage{dcolumn}
\usepackage{bm}
\usepackage{color}
\usepackage{txfonts}
\usepackage{microtype}
\usepackage{overpic}

\begin{document}
\author{Yan-Fei Li}	\affiliation{School of Science, Xi'an Jiaotong University, Xi'an 710049, China}
\author{Yong-Tao Zhao}
\affiliation{School of Science, Xi'an Jiaotong University, Xi'an 710049, China}
\author{Karen Z. Hatsagortsyan}
\affiliation{Max-Planck-Institut f\"{u}r Kernphysik, Saupfercheckweg 1,
	69117 Heidelberg, Germany}
\author{Christoph H. Keitel}
\affiliation{Max-Planck-Institut f\"{u}r Kernphysik, Saupfercheckweg 1,
	69117 Heidelberg, Germany}	
\author{Jian-Xing Li}\email{jianxing@xjtu.edu.cn}
\affiliation{School of Science, Xi'an Jiaotong University, Xi'an 710049, China}
\bibliographystyle{apsrev4-1}

\title{Electron-Angular-Distribution Reshaping  in Quantum Radiation-Dominated Regime }

\date{\today}

\begin{abstract}

Dynamics of an electron beam head-on colliding  with an ultraintense focused ultrashort circularly-polarized laser pulse are investigated in the quantum radiation-dominated regime.
Generally, the ponderomotive force of the laser fields may deflect  the electrons transversely, to form a ring structure in the cross-section of the electron beam. However, we find that when the Lorentz factor of the electron $\gamma$ is approximately one order of magnitude larger than the invariant laser field parameter $\xi$, the stochastic nature of the photon emission
leads to  electron aggregation abnormally inwards
to the propagation axis of the laser pulse. Consequently, the electron angular distribution after the interaction exhibits a peak structure in the beam propagation direction, which is apparently distinguished from the ``ring''-structure of the
distribution  in the classical regime, and therefore, can be recognized as a proof of the fundamental quantum stochastic nature of radiation.
The  stochasticity signature is robust with respect to the laser and electron parameters and  observable with current experimental techniques.

\end{abstract}

\maketitle

\section{introduction}
Rapid development of ultrashort ultraintense laser techniques \cite{Yanovsky2008,Danson2015,Sung2017,Vulcan,ELI,Exawatt} has significantly stimulated the worldwide research interests not only on  novel applications of laser-matter interaction \citep{Leemans2006GeV,Schwoerer2006Laser,Hegelich2006,Sarri2014}, but also on the investigation of fundamental issues \citep{Piazza2012,Esarey2009,Mourou2006,Marklund2006}.  An example is radiation reaction (RR), which has been discussed since the early days of classical and quantum electrodynamics \cite{Abraham_1905,Lorentz_1909,Dirac_1938,Heitler_1941}, with the testing of the theory being  experimentally realized only recently \cite{Cole_2018,Poder_2018}.
In ultrastrong laser fields the radiative processes may reach the quantum regime \cite{Goldman_1964,Nikishov_1964,Ritus_1985,Sokolov2009,Sokolov2010,Piazza2010,Blackburn_2014,Li2014,
Li2015,Dinu2016,Vranic2016,Harvey2017,Piazza2013,Piazza2014,Li2017}. One of the  significant quantum properties of radiation is the stochastic nature, i.e., the discrete and probabilistic character of photon emission \cite{Piazza2013,Piazza2014,Tamburini_2014,Bashinov2015,Wang2015,Li2017}. One signature of stochasticity effects (SE) of radiation  is the so-called electron straggling effect, which results in quantitative increase of the yield of the high-energy photons in strong fields\cite{Blackburn_2014}, and the quantum quenching of radiation losses in subcycle petawatt lasers \cite{Harvey2017}.
Theoretically it has also been shown that the SE can  broaden  the energy spread of the electron beam in a plane laser field \cite{Piazza2013,Piazza2014} and cause electron stochastic heating in a standing wave \cite{Bashinov2015}.
In a focused laser pulse the SE modified by the ponderomotive force may produce an additional energy spread, as for instance, has been shown in
\cite{Wang2015}. 
  Compared with radiative SE signatures  \cite{Blackburn_2014, Harvey2017, Li2017}, the relevant signatures in the electron dynamics
  may be easier for experimental observation, since the diffraction limitation of an electron is much smaller than that of a photon. In this paper we aim at to identify such SE signature in electron dynamics, which would have a qualitative nature and, consequently, would be straightforwardly distinguishable at current achievable experimental conditions.

The invariant parameter that characterizes quantum effects in the strong field processes is $\chi\equiv |e|\hbar\sqrt{(F_{\mu\nu}p^{\nu})^2}/m^3c^4$ \cite{Ritus_1985}, where  $F_{\mu\nu}$ is the field tensor, $\hbar$  the reduced Planck constant, $c$ the speed of the light in vacuum, $p^{\nu}=(\varepsilon/c,\textbf{p})$  the incoming electron 4-momentum, and $-e$ and $m$ are the electron charge and mass, respectively.  When the electron counterpropagates with the laser beam, one may estimate $\chi\approx 2(\hbar\omega_0/mc^2)\xi\gamma$. Here, $\xi\equiv |e|E_0/(m\omega_0 c)$  is the invariant laser field parameter, $E_0$ and $\omega_0$ are the amplitude and frequency of the laser field, respectively, and $\gamma$ is the electron Lorentz-factor. SE are expected to be large when RR is significant, i.e., in the quantum radiation-dominated regime (QRDR), which requires $R\equiv \alpha\xi\chi\gtrsim 1$ \cite{Piazza2012,Koga2005}, indicating that the radiation losses during a laser period are comparable with the electron initial energy. $\alpha$ is the fine structure constant.
With the worldwide construction of petawatt laser facilities, laser pulses with an intensity above $10^{22}$ W/cm$^2$ ($\xi\sim 10^2$) are available nowadays \cite{Danson2015,Yanovsky2008,Sung2017}, and much more intense lasers will be produced in the near future \cite{Vulcan,ELI,Exawatt}. Meanwhile, the energies of electrons accelerated by a laser wakefield can be up to several GeV ($\gamma\sim 10^3$) \cite{Mourou2006,Esarey2009,Leemans2006GeV}. Thus, the conditions for SE measurement, $\chi\sim 1$ and $R\sim 1$, are achievable with current experimental techniques. Recently, innovative experimental evidences of quantum RR effects have been realized through radiation  spectra from ultrarelativistic positrons in silicon in a regime where quantum RR effects dominate the
positron dynamics \cite{Wistisen2018Experimental}, and through the electron energy loss in all-optical experiments \cite{Cole_2018,Poder_2018}, respectively. However, in those experiments all quantum properties, including SE and photon recoil effect, arise simultaneouly, rendering it challenging to identify SE in combination with an appropriate  set of theoretical methods.

  \begin{figure}[t]
	\includegraphics[width=1\linewidth]{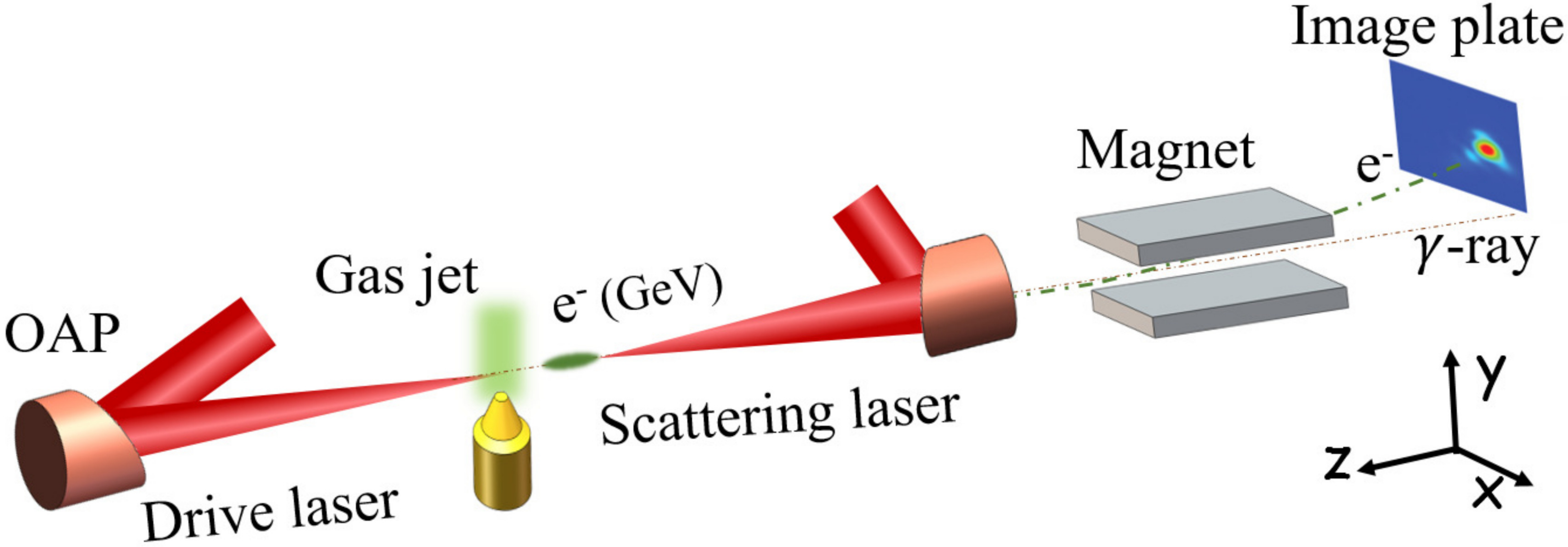}
	\caption{Scenario of SE observation in electron-beam dynamics in nonlinear Compton scattering. An electron beam with a mean kinetic energy of about GeV generated by the laser wakefield accelerator head-on collides with an ultraintense scattering laser pulse.  The electrons aggregate inwards to the propagation axis of the laser pulse due to the stochastic nature of the photon emission, which can be observed by the electron angular distribution at the image plate. A magnet is required to split the electron beam from the $\gamma$-ray radiation.   }
	\label{fig1}
\end{figure}

In this paper, we investigate the SE of  photon emissions on the electron-beam dynamics in QRDR, see the interaction scheme in Fig.~\ref{fig1}. A GeV electron beam generated by the laser wakefield acceleration head-on collides with an ultraintense laser pulse. We consider the parameter conditions  $\chi\sim 1$ and $\gamma$ is approximately one order of magnitude larger than $\xi$: the former ensures the SE being significant and dominating the electron-beam dynamics, and the latter facilitates the SE under observation, namely, the electron-beam aggregation effect at the center of the electron angular distribution, which overcomes the electron-beam expansion  produced by the ponderomotive force due to the transverse profile of the focused laser fields.
The electron-beam aggregation effect produces a peak in  the electron angular distribution with a FWHM  larger than about $40^{\circ}$, which is robust with respect to the laser and electron parameters for current achievable experimental techniques.

This paper is organized as follows. In Sec. II
we discuss  the applied theoretical models for the calculation of the electron dynamics and radiation. In Sec. III th SE signature in electron angular distribution is represented and analyzed.  
In Sec. IV we investigate the impacts of the laser and electron parameters on the SE signature. A brief conclusion 
of our work is given in Sec. V.

\section{Applied theoretical models for the calculation of the electron dynamics and radiation}

We use a theoretical model to calculate  the electron dynamics based on Monte-Carlo simulations employing QED theory for the electron radiation and classical equations of motion for the propagation of electrons between photon emissions \cite{Elkina2011,Ridgers2014,Green2015}, which is indicated as Monte-Carlo model (MCM). In ultraintense laser fields,  $\xi \gg1$, the coherence length of the photon emission is much smaller than the laser wavelength and the typical size of the electron trajectory \cite{Ritus1985,Khokonov2010}. As a result, the photon emission probability is determined by the local electron trajectory, consequently, by the local value of the parameter $\chi$ \cite{Baier1994}. In every step (far less than the coherence length of the photon emission), the emission process is implemented as a random process, see below. In MCM, the quantum properties of SE and photon recoil are included. In order to highlight the impacts of SE, we carry out additional calculations excluding SE but including other key quantum effects. The latter is based on the, so-called, modified Landau-Lifshitz equation. Generally, the Landau-Lifshitz equation \cite{Landau1975} describes electron dynamics under the action of RR in the classical regime $\chi\ll 1$.
In the case of $\chi\sim 1$, the classical Landau-Lifshitz model (LLM) overestimates the RR force, which is remedied phenomenologically
in the modified Landau-Lifshitz model (MLLM)
\cite{Poder_2018,Piazza2012}. Note that the latter provides results similar to
the semi-classical Sokolov equation  \cite{Sokolov2009,Sokolov2010}.
The MLLM treats electron dynamics classically taking into account the quantum-recoil in RR, however, neglecting SE in photon emission. 

For easily comprehending our simulation results, the three models are briefly introduced as follows.

\subsection{Landau-Lifshitz Model}
In this model, the RR is considered as the effect of the elec-
tromagnetic fields emitted by an electron on the motion of
itself classically. The dynamics of an electron is discribed by the Landau-Lifshitz (LL) equation \cite{Landau1975}
\begin{equation}
 m \frac{du^{\mu}}{d\tilde{\tau}}=eF^{\mu j}u_j + f^{\mu},
\end{equation}
 where
\begin{equation}
f^{\mu}=\frac{2e^3}{3mc^2}(\partial_\alpha F^{\mu \nu}u_\nu u^\alpha)+\frac{2e^4}{3m^2c^4}(F^{\mu \nu}F_{\nu \mu}\mu^\alpha+(F^{\nu \beta}u_\beta F_{\nu \alpha}u^\alpha)u^\mu),
\end{equation}
$u^\mu=(\gamma,\gamma \textbf{v}/c)$ is four-velocity of the electron, $\tilde{\tau}$ the proper time,
\begin{equation}
\frac{d}{d\tilde{\tau}}=(k\cdot p)\frac{d}{d\tilde{\eta}}, \quad \tilde{\eta}=(k\cdot \tilde{r}),
\end{equation} 
and $\tilde{r}$ the four-vector of the coordinate.
The three-dimention equation is 
\begin{widetext}
 \begin{eqnarray}\label{LL}
{\bm F_{RR, classical}}&=&\frac{2e^3}{3mc^3} \left(\gamma\left(\left(\frac{\partial}{\partial t}+\frac{\textbf{p}}{\gamma m}\cdot\nabla\right){\textbf{E}}+\frac{{\textbf{p}}}{\gamma m c}\times\left(\frac{\partial}{\partial t}+\frac{{\textbf{p}}}{\gamma m}\cdot\nabla\right){\textbf{B}}\right)\right.\nonumber\\
&& +\frac{e}{m c}\left({\textbf{E}}\times{\textbf{B}}+\frac{1}{\gamma m c}{\textbf{B}}\times\left({\textbf{B}}\times{\textbf{p}}\right)+\frac{1}{\gamma m c}{\textbf{E}}\left({\textbf{p}}\cdot{\textbf{E}}\right)\right)\nonumber\\
&& \left. -\frac{e\gamma}{m^2 c^2}{\textbf{p}}\left(\left({\textbf{E}}+\frac{{\textbf{p}}}{\gamma m c}\times {\textbf{B}}\right)^2-\frac{1}{\gamma^2m^2c^2}\left({\textbf{E}}\cdot{\textbf{p}}\right)^2\right)\right),
\end{eqnarray}
\end{widetext}
where $\textbf{E}$ and $\textbf{B}$ are the electric and magnetic fields, respectively.

\subsection{Modified Landau-Lifshitz Model}
In this model we treat the electron dynamics in the external
field classically but take into account the quantum-recoil
corrections. The equation used to calculate the electron dynamics is the modified-LL equation with the classical RR force in the LL equation replaced by the quantum RR force as \cite{Poder_2018,Piazza2012}:
\begin{equation}
{\bm F_{RR, quantum}}=\frac{I_{QED}}{I_{C}}{\bm F_{RR, classical}},
\end{equation}
where,
\begin{eqnarray}
I_{QED}&=&mc^2\int c\left(k\cdot k' \right)\frac{d W_{fi}}{d\eta dr_0}dr_0,\\
I_{C}&=&\frac{2e^4E'^2}{3m^2c^3},
\end{eqnarray}
$W_{fi}$ is the radiation probability, $r_0=\frac{2\left(k\cdot k'\right)}{3\chi\left(k\cdot p_i\right)}$, and $E'$ is the electric fields in the electron frame. $k$, $k'$ and $p_i$ are the four-vector of the wave vector of the driving laser, the wave vector of the radiated photon, and the momentum of the electron before the radiation, respectively.

In the modified-LL equation, the recoil effects are included by renormalizing the RR force by the factor $I_{QED}/I_{C}$, the ratio of the radiation intensities within QED and classical approaches, which will account for the classical overestimation of the RR effects on electron dynamics.

Note that the same results as the modified-LL equation are provided by the phenomenologically derived equation of motion for an electron in the $\xi \gg 1$ limit,  based on the energy-momentum conservation within the system of the electron and emitted photons at each formation length of radiation \cite{Sokolov2009,Sokolov2010}:

\begin{eqnarray}
\frac{d p^{\alpha}}{d\tilde{\tau}}=\frac{e}{m c}F^{\alpha\beta}p_\beta-\frac{{\cal I}_{QED}}{m c^2}p^{\alpha}+\tau_c\frac{e^2{\cal I}_{QED}}{m^2c^2{\cal I}_c}F^{\alpha\beta}F_{ \beta\gamma}p^{\gamma},
\end{eqnarray}

where  $\tau_c\equiv 2e^2/(3mc^3)$.

\subsection{Monte-Carlo Model}
In this model, the calculation of the electron dynamics is based on the Monte-Carlo simulations employing QED theory for the electron radiation and classical equations of motion for the propagation of electrons between photon emission  \cite{Elkina2011,Ridgers2014,Green2015}.

In superstrong laser fields $\xi\gg 1$, the photon emission
 probability $W_{fi}$ is determined by the local electron trajectory, consequently, by the local value of the parameter $\chi$ \cite{Ritus1985}:
\begin{eqnarray}
\frac{d^2 W_{fi}}{d \tilde{\eta} dr_0}=\frac{\alpha \chi [\int_{r_{\chi}}^{\infty} K_{5/3}(x)dx+r_0 r_{\chi} \chi^2 K_{2/3}(r_{\chi})]}{\sqrt{3}\pi\lambdabar_c(k\cdot p_i)},
\label{W}
\end{eqnarray}
where the Compton wavelength $\lambdabar_c=\hbar/mc$, and $r_{\chi}  =r_0/(1-3\chi r_0/2)$.
The photon emission of electrons is considered to be a Monte-Carlo stochastic process \cite{Elkina2011,Ridgers2014,Green2015}. During the electron-laser interaction, for each propagation coherent length $\Delta \tilde{\eta}$, the photon emission will take place if the condition $(dW_{fi}$/d$\tilde{\eta})\Delta\tilde{\eta}\geq N_{r}$ is fulfilled, where $N_r$ is a uniformly distributed random number in $[0, 1]$. Herein, the coherent length $\Delta \tilde{\eta}$ is inversely proportional to the invariant laser field parameter $\xi$, i.e., $\Delta \tilde{\eta}\sim 1/\xi$. However, to keep the total photon emission energy consistent, i.e., to exclude numerical error of the simulation of photon emission, we choose $\Delta \tilde{\eta} \ll 1/\xi$. The photon emission probability
\begin{eqnarray}
W_{fi} = \Delta \tilde{\eta}\frac{dW_{fi}}{d\tilde{\eta}}=\Delta \tilde{\eta}\int_{ \omega_{min}}^{ \omega_{max}} \frac{d^2 W_{fi}}{d \tilde{\eta} d\omega}d\omega, \nonumber
\end{eqnarray}
where $\omega_{min}$ and $\omega_{max}$ are assumed to equal the driving laser photon energy and  the electron instantaneously kinetic energy, respectively.
In addition, the emitted photon energy $\omega_R$ is determined by the relation:
\begin{eqnarray}
\frac{1}{W_{fi}}\int_{\omega_{min}}^{\omega_R}\frac{d W_{fi}(\omega)}{d\omega}d\omega = \frac{\Delta\tilde{\eta}}{W_{fi}}\int_{\omega_{min}}^{\omega_R}\frac{d^2 W_{fi}(\omega)}{d\tilde{\eta} d\omega}d\omega=\tilde{N}_r,\nonumber
\end{eqnarray}
where, $\tilde{N}_r$ is another independent uniformly distributed random number in $[0, 1]$.
 Between the photon emissions, the electron dynamics in the laser field is governed by classical equations of motion:
\begin{eqnarray}
\frac{d\bf{p}}{dt}=e({\bf E}+\frac{\textbf{v}}{c}\times\textbf{B}).
\end{eqnarray}
Given the smallness of the emission angle $ \sim 1/\gamma$  for an ultrarelativistic electron, the photon emission is assumed to be along the electron velocity. The photon emission induces the electron momentum change ${\bf p}_f \approx (1-\omega_R/c|{\bf p}|) {\bf p}_i$, where ${\bf p}_{i,f}$ are the electron momentum before and after the emission, respectively.

\subsection{Employed electromagnetic fields of the laser pulse}

In this work, we employ a circularly-polarized tightly-focused laser pulse with a Gaussian temporal profile, which propagates along +z-direction as a scattering laser beam. The spatial distribution of the electromagnetic fields takes into account up to $\epsilon^3$-order of the nonparaxial solution \cite{Salamin2002,Salamin2002PhysRevSTAB}, where $\epsilon=w_0/z_r$, while $w_0$ is the laser beam waist, $z_r=k_0w_0^2/2$ the Rayleigh length with laser wave vector $k_0=2\pi/\lambda_0$, and $\lambda_0$ the laser wavelength. The expressions of the electromagnetic fields are presented in the following  \cite{Salamin2002,Salamin2002PhysRevSTAB}:

\begin{eqnarray}
E_x &=& -i E\left[1+\epsilon^2\left(f^2\widetilde{x}^2-\frac{f^3\rho^4}{4} \right) \right],\\
E_y &=& -i E \epsilon^2 f^2 \widetilde{x}\widetilde{y},\\
E_z &=& E\left[\epsilon f \widetilde{x} + \epsilon^3 \widetilde{x} \left(-\frac{f^2}{2}+f^3\rho^2-\frac{f^4\rho^4}{4}\right) \right],\\
B_x &=& 0,\\
B_y &=& -i E \left[1+\epsilon^2\left(\frac{f^2\rho^2}{2}-\frac{f^3\rho^4}{4} \right)\right],\\
B_z &=& E\left[\epsilon f \widetilde{y} + \epsilon^3 \widetilde{y} \left(\frac{f^2}{2}+\frac{f^3\rho^2}{2}-\frac{f^4\rho^4}{4}\right) \right],
\end{eqnarray}
where,
\begin{equation}
E = E_0 F_n f e^{-f\rho^2} e^{i\left(\eta+\psi_{\rm{CEP}}\right)} e^{-\frac{t^2}{\tau^2}},
\end{equation}
$\tau$ is the laser pulse duration, and $E_0$ the amplitude of the laser fields with normalization factor $F_n=i$ to keep $\sqrt{E_x^2+E_y^2+E_z^2}=E_0$ at the focus, yielding the scaled coordinates
\begin{equation}
\widetilde{x}=\frac{x}{w_0}, \quad \widetilde{y}=\frac{y}{w_0},\quad \widetilde{z}=\frac{z}{z_r}, \quad \rho^2=\widetilde{x}^2+\widetilde{y}^2,
\end{equation}
where $f=\frac{i}{\widetilde{z}+i}$, $\eta=\omega_0 t-k_0z$, and $\psi_{\rm{CEP}}$ is the carrier-envelope phase.

\section{The SE signature in electron angular distribution}

 \begin{figure}[t]
	\includegraphics[width=1.0\linewidth]{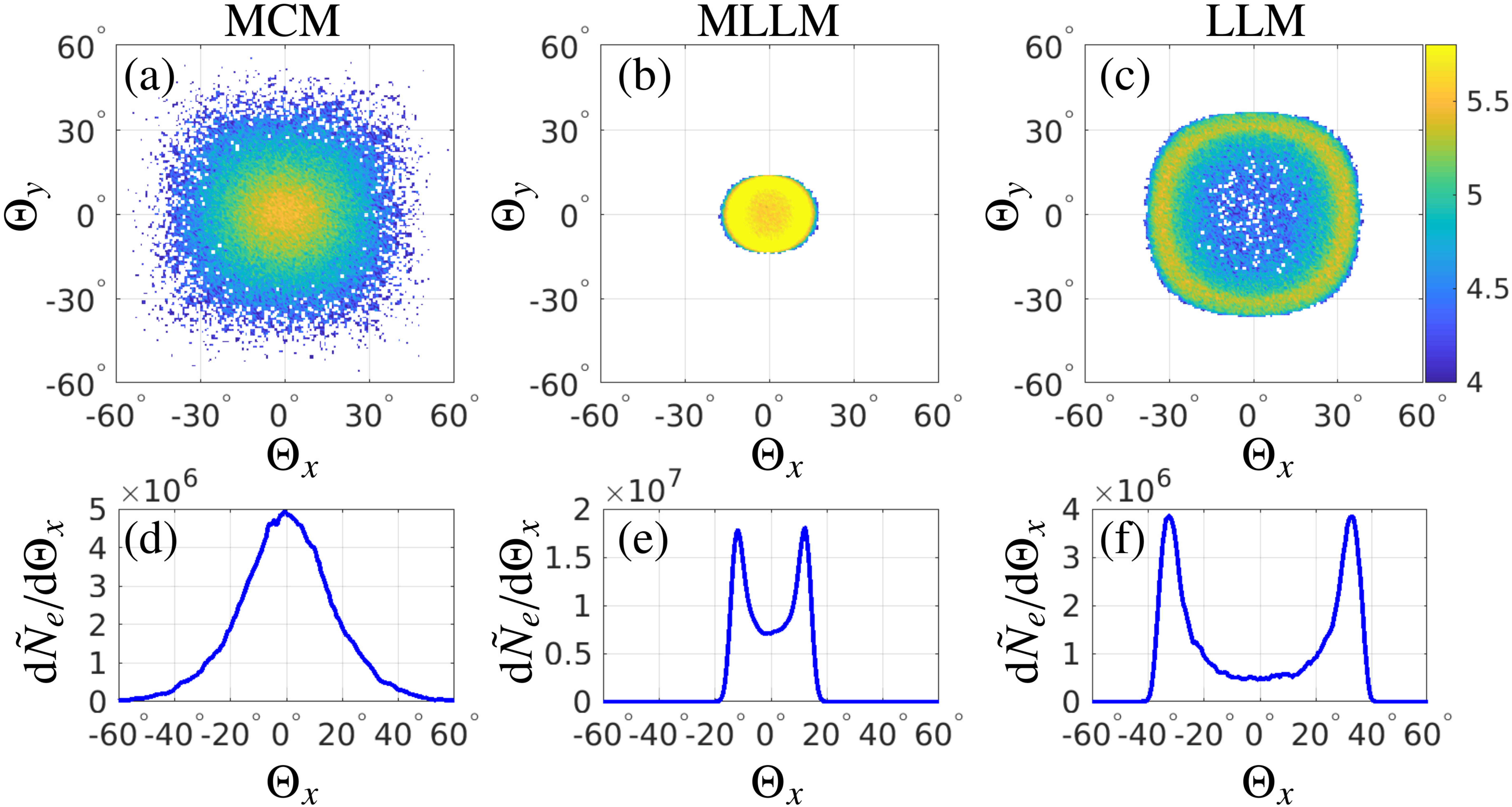}
	\caption{Electron angular distribution: (a)-(c)
		log$_{10}$[d$^2N_e$/(d$\Theta_x$d$\Theta_y$)] rad$^{-2}$  vs the transverse deflection angles of the electron momenta  $\Theta_x \equiv$ arctan($p_x/p_z$) and $\Theta_y \equiv$ arctan($p_y/p_z$). The color bar in (c) applies for (a) and (b) as well.
		(d)-(f):		The angular distribution integrated
	over the angle region $-5^{\circ}\leq\Theta_y\leq5^{\circ}$, d$\tilde{N}_e$/d$\Theta_x$ = $\int_{-5^{\circ}}^{5^{\circ}}$d$^2N_e$/[d$\Theta_x$d$\Theta_y$] d$\Theta_y$,  vs $\Theta_x$ corresponding to simulations of (a)-(c), respectively. Simulations are calculated via MCM ((a) and (d)), MLLM ((b) and (e)), and LLM ((c) and (f)), respectively. 
		The parameters of the laser and electron beam are given in the text.}
	\label{fig2}
\end{figure}

An electron beam with characteristics like via laser wakefield accelerators is employed to interact with a counterpropagating focused laser pulse in QRDR, see Fig.~\ref{fig1}. We consider the interaction regime  $\gamma\gg\xi/2$, when even in QRDR the electron  forward motion persists 
and the deflection angle in the transverse plane observed on the image plate is mostly determined by the ponderomotive potential due to the transverse profile of the laser beam. We may estimate the deflection angle as $\theta_{d}\sim |F_{p\perp}|\tau/p_{\parallel}\propto (\xi^2/\gamma^2)(\tau/w_0)$, with relativistic ponderomotive force $\textbf{F}_p = -\nabla \xi^2/(2\gamma)$ \cite{Quesnel_1998}, and laser pulse duration $\tau$.
This is in contrast to the, so-called, reflection regime $\gamma\lesssim \xi/2$, when the electron is reflected backwards with respect to its initial motion because of combined action of RR and laser ponderomotive force \citep{Salamin1996,Li2015,Li2018}.

We investigate the electron dynamics by employing MCM, MLLM, and LLM, respectively, and the corresponding angle-resolved electron-number distributions  with respect to the transverse deflection angles of the electron momenta are illustrated in Fig.~\ref{fig2}.  The laser peak intensity  $I_0\approx 1.4\times10^{22}$ W/cm$^2$ ($\xi=100$), $\lambda_0=1 \mu$m, $w_0=4 \mu$m, the FWHM of laser pulse duration $\tau=16 T_0$, and $T_0$ is the laser period. The pair production probility in such ultrashort laser pulse is negligible. The initial mean kinetic energy of electrons is $\varepsilon_i=1$ GeV ($\gamma\approx 1956.95$, $R\approx 1$, and $\chi_{max}\approx 1.38$) with an energy spread $\Delta\varepsilon_i/\varepsilon_i=0.02$.  A cylindrical electron beam is employed, and the beam parameters are set as: radius $w_e= 2 \lambda_0$, length $L_e = 8 \lambda_0$, angular spread $\Delta \theta_e\approx\pm3.6^\circ$, and electron number $N_e = 1.5\times10^5$ (i.e., density $n_e\approx 10^{15}$ cm$^{-3}$ with a Gaussian density profile on the cross section of the electron beam). Those electron-beam parameters are achievable for current laser-plasma acceleration setups \citep{Leemans2014,Kneip2009,Clayton2010,Pollock2011}. 

   \begin{figure}[t]
  	\includegraphics[width=0.95\linewidth]{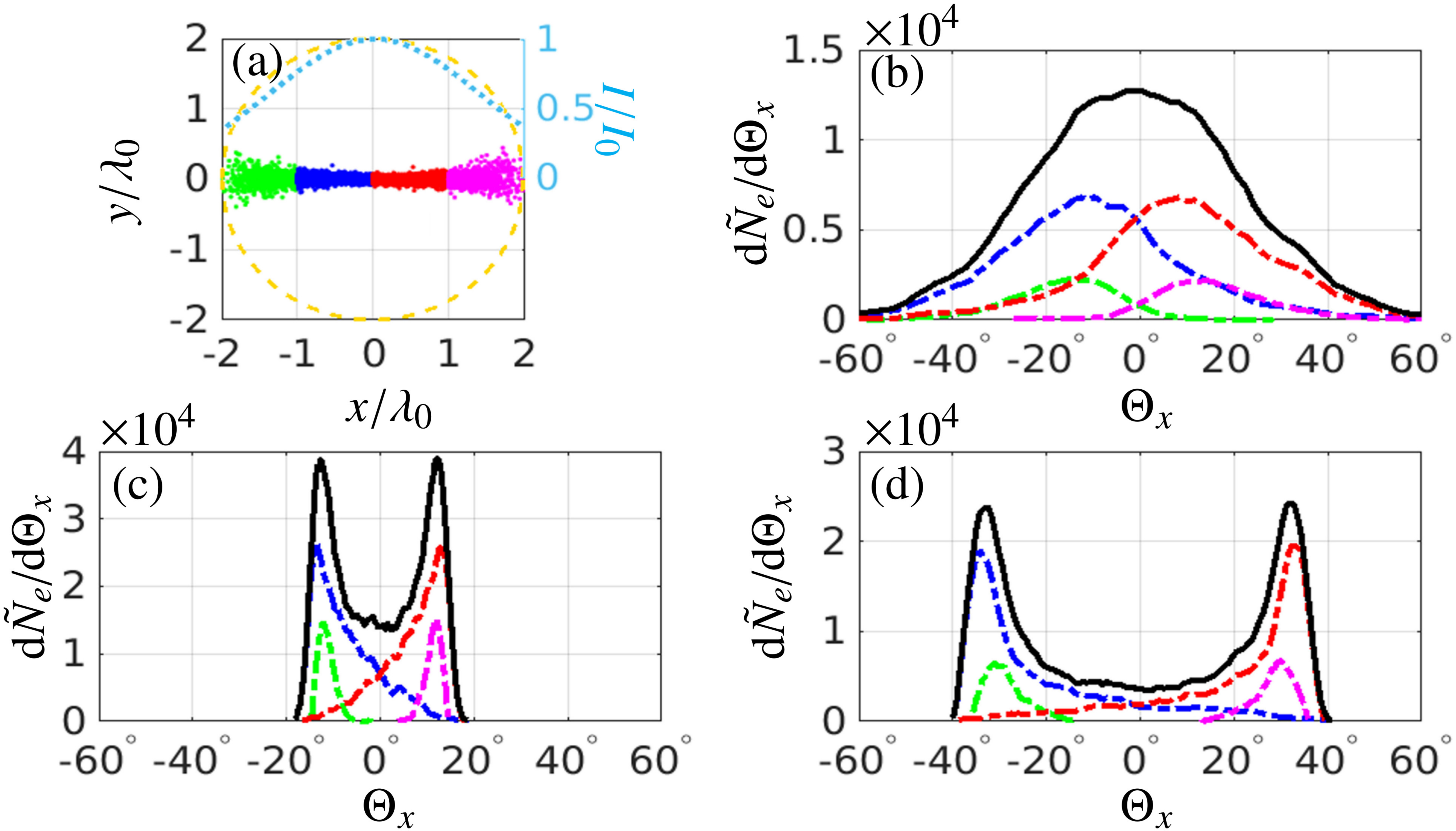}
  	\caption{(a) The initial transverse coordinate distribution of electrons near the $x-z$ plane at $y=0$, which finally contribute to the angular distribution peaks. The cyan-dotted curve in (a) shows the transverse profile of the laser intensity $I$ scaled by $I_0$.
     The yellow circle represents the boundary of the electron beam, and different colors show the different sample electrons.  (b)-(d) Electron angular distribution
      after the interaction via  MCM,  MLLM and LLM, respectively. The dash-dotted curves of different colors represent the electron distributions of different sample electrons indicated
        in (a).  The black-solid curves indicate the total electron angular distribution. Other laser and electron beam parameters are the same as in Fig.~\ref{fig2}. }
  	\label{fig3}
  \end{figure}

The MCM simulation which includes SE in Fig.~\ref{fig2}(a) shows that the electrons move inwards to the propagation axis of the laser pulse, consequently, a broad electron-density peak emerges
in the middle of the electron  angular distribution,
which decays exponentially to the peripheries. The electrons concentrate with an angular radius of about $40^\circ$.
When  SE are excluded, as in MLLM and LLM simulations, see Figs.~\ref{fig2}(b)  and \ref{fig2}(c), respectively, the electron angular distributions in both cases have a ``ring'' structure, and the density decays exponentially inwards to the center and outwards to the peripheries. This is because the ponderomotive force deflects the electrons transversely outwards.
The SE overcome the deflection effect of the ponderomotive force and cause electrons aggregation inwards to the laser propagation axis.
The angular radius of the density ``ring'' is approximately $20^{\circ}$ for MLLM, but $40^{\circ}$ for LLM, since in the latter the LL equation overestimates the RR force, and in the deflection angle estimation $\gamma$ should be replaced by $(\varepsilon_i-\varepsilon_R)/m $,
where $\varepsilon_R$ is the electron energy loss due to the radiation.
For a quantitative analysis we integrate the electron differential angular distributions in the angular range of $-5^\circ \lesssim \Theta_y \lesssim 5^\circ$, which are represented in Figs.~\ref{fig2}(d)-\ref{fig2}(f), respectively.
For MCM, MLLM, and LLM, the electron-density peaks are at $\Theta_x=0^\circ$, $\pm 12^\circ$, and $\pm 33^\circ$, respectively, and the corresponding FWHMs are about $34^\circ$, $7^\circ$, and $12^\circ$, respectively.
The current techniques of electron detectors with an angular resolution less than 0.1 mrad \cite{Wang2013,Leemans2014,Wolter2016,Chatelain2014} will allow to experimentally distinguish  the angular distributions of the MCM case  with those via MLLM and LLM, and in this way identify the SE role. Since $\varepsilon_i$ is too large at chosen parameters,
the observation of the electron-number distribution is more convenient than that of the electron-energy distribution  \cite{Wang2013,Leemans2014,Wolter2016,Chatelain2014}.

 To analyze the role of SE in
  forming the electron distribution, we follow the tracks of a group of sample electrons near the $x-z$ plane at $y=0$, see Fig.~\ref{fig3}. The initial coordinate distribution of the sample electrons are shown in Fig.~\ref{fig3}(a). Note that the electron density has a transverse Gaussian distribution in the cross section of the electron beam, such that the numbers of electrons marked in blue and red are
  larger than those in green and magenta. Under the deflection effects of the laser fields, electrons in different groups      (marked in different colors)
   produce different profile curves in
  the final angular
    distributions in Figs.~\ref{fig3}(b)-\ref{fig3}(d). Apparently, as SE are excluded, see Figs.~\ref{fig3}(c) and \ref{fig3}(d), the sample electrons mainly move outwards under the transverse ponderomotive force. Since $w_0=2w_e$ and the laser-intensity gradient near the peak is small, see Fig.~\ref{fig3}(a), the electrons experience similar laser fields, and consequently, the deflection angle $\theta_d$ concentrates at $\Theta_x=\pm12^\circ$ and $33^\circ$, respectively, with small angular spreads, which is proportional to the laser intensity gradient.
Finally, a  ``ring'' structure emerges in the electron angular distributions, see Figs.~\ref{fig2}(b) and \ref{fig2}(c).
As the SE are necessarily taken into account, comparing Fig.~\ref{fig3}(b) with Fig.~\ref{fig3}(c), the SE in photon emission induce  stochastic electron dynamics, and consequently a large spread of the final electron momenta.
All electrons have substantial probabilities of moving inwards to the laser propagation axis,
which leads to the overlap of angular distributions from different electron groups and the formation of the electron-density peak at $\Theta_x=0^\circ$.

\begin{figure}[t]
	\includegraphics[width=1.0\linewidth]{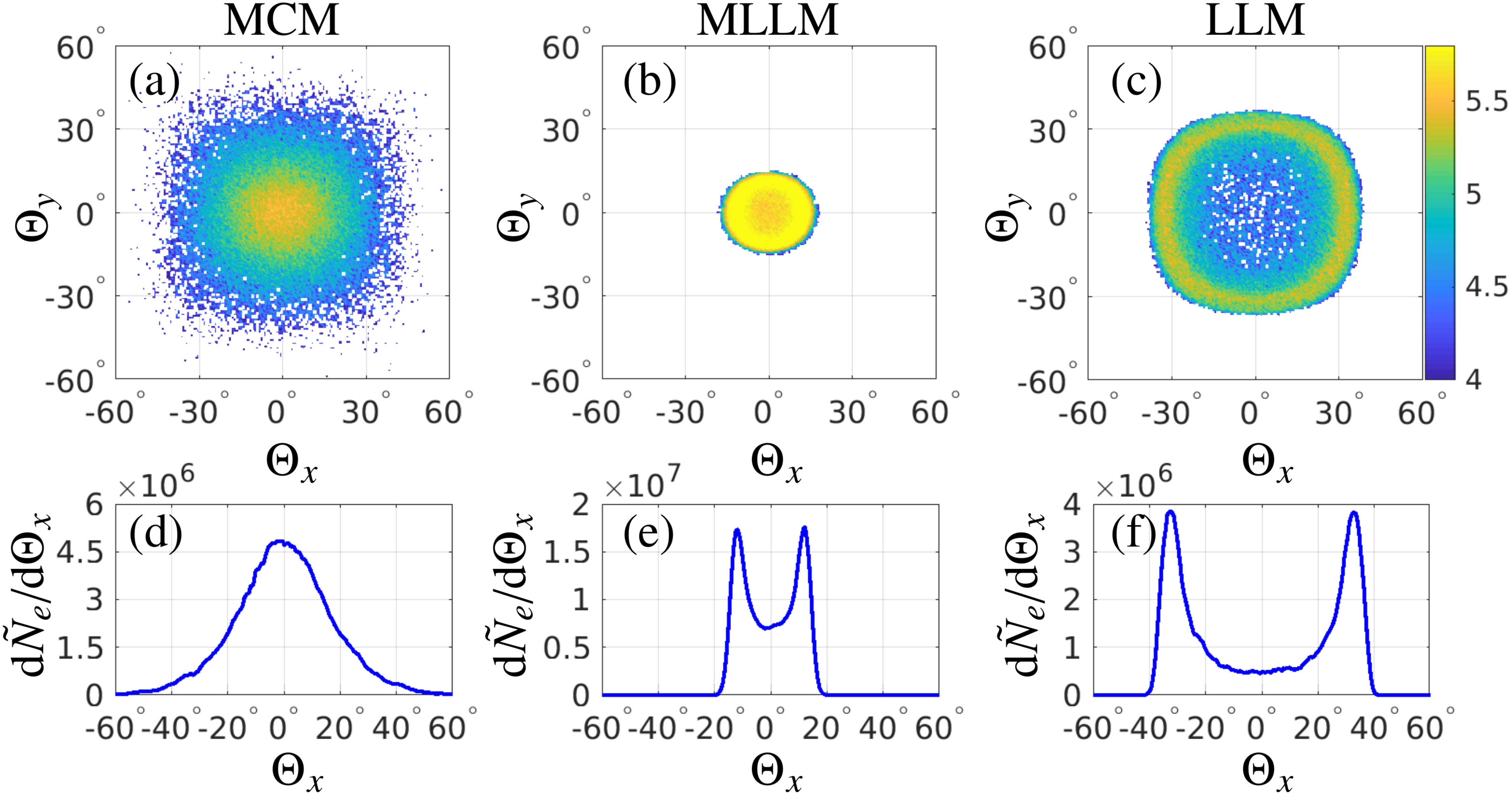}
	\caption{(a)-(c): Electron angular distribution after the interaction,  
log$_{10}$[d$^2N_e$/(d$\Theta_x$d$\Theta_y$)] rad$^{-2}$  vs  $\Theta_x$ and $\Theta_y$.  The color bar in (c) applies for (a) and (b) as well.
(d)-(f): 
d$\tilde{N}_e$/d$\Theta_x$  with respect to $\Theta_x$ corresponding to (a)-(c), respectively.  The electron dynamics are simulated via ((a) and (d)) MCM, ((b) and (e)) MLLM, and ((c) and (f)) LLM, respectively. $\Delta\varepsilon_i/\varepsilon_i=0.1$, and other laser and electron parameters are the same as in Fig.~\ref{fig2}.}
	\label{fig4}
\end{figure}

\begin{figure}[t]
	\includegraphics[width=1.0\linewidth]{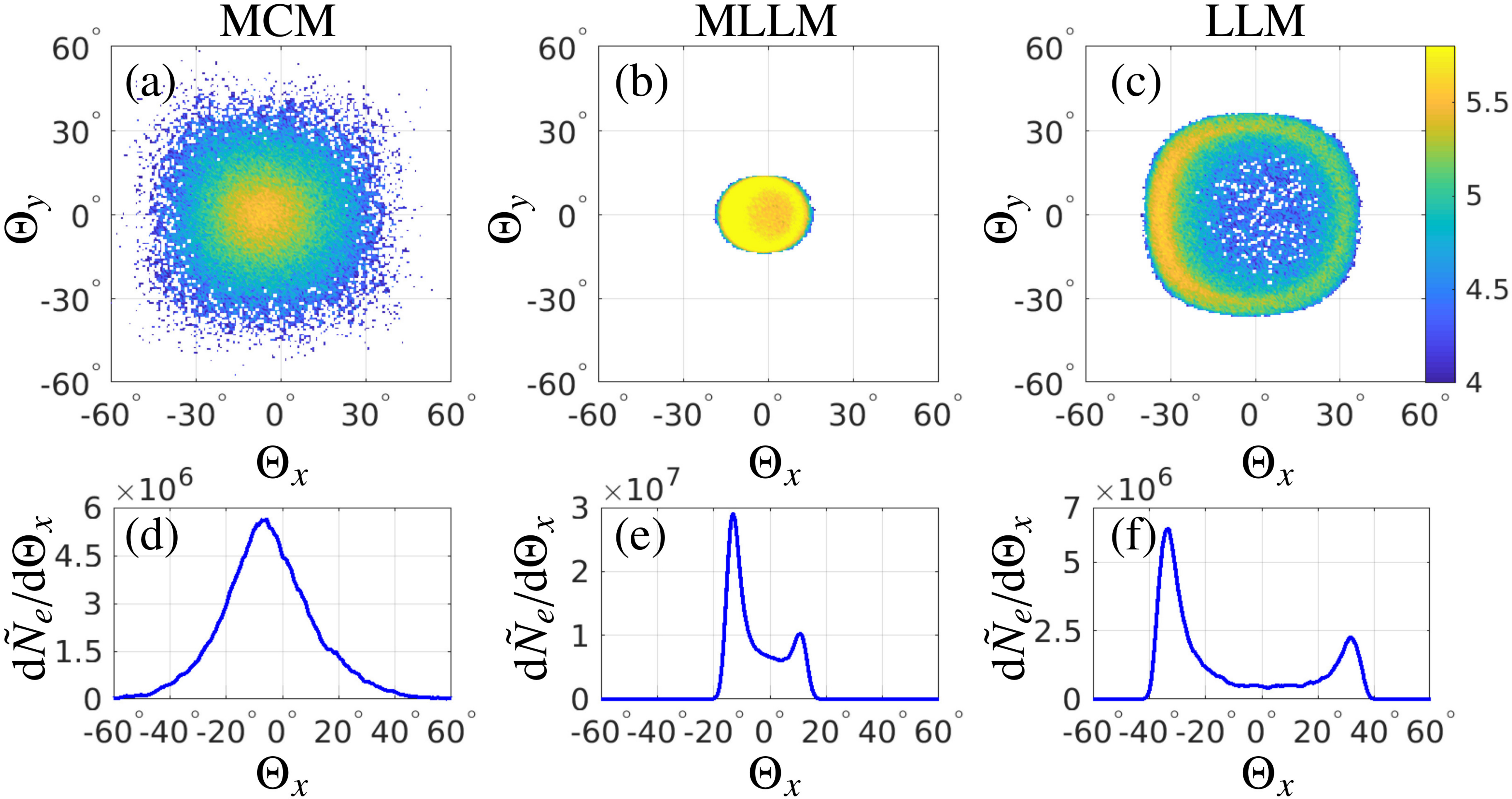}
	\caption{(a)-(c): Electron angular distribution after the interaction 
log$_{10}$[d$^2N_e$/(d$\Theta_x$d$\Theta_y$)] rad$^{-2}$  vs $\Theta_x$ and $\Theta_y$.  The color bar in (c) applies for (a) and (b) as well.
(d)-(f): 
d$\tilde{N}_e$/d$\Theta_x$  vs $\Theta_x$ corresponding to (a)-(c), respectively. The electron dynamics are simulated via ((a) and (d)) MCM, ((b) and (e)) MLLM, and ((c) and (f)) LLM, respectively. The collision angle of the electron beam $\theta_e=179^{\circ}$, and other laser and electron parameters are the same as in Fig.~\ref{fig2}.}
	\label{fig5}
\end{figure}

 \begin{figure}[b]
	\includegraphics[width=1.0\linewidth]{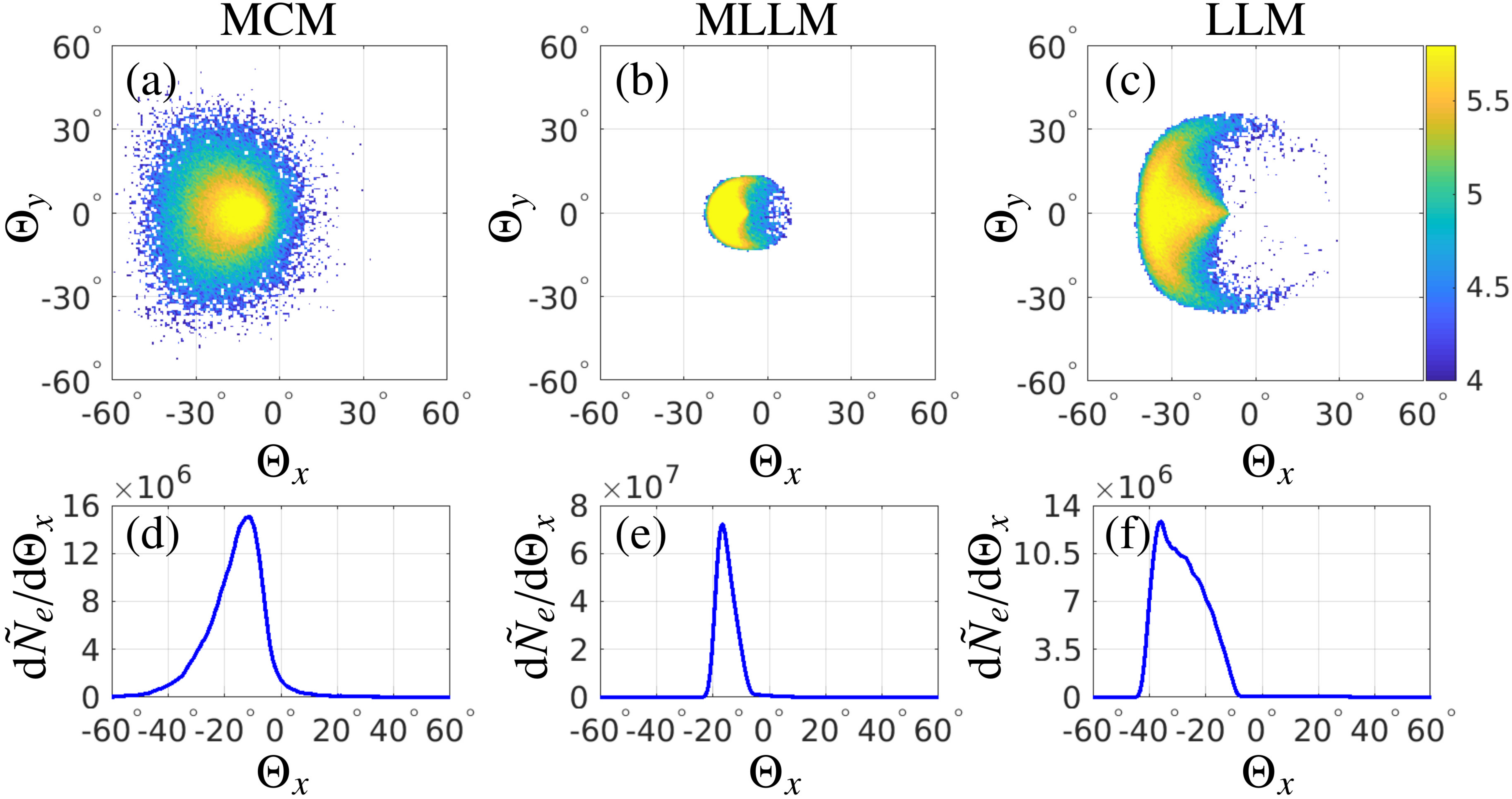}
	\caption{(a)-(c): Electron angular distribution after the interaction 
log$_{10}$[d$^2N_e$/(d$\Theta_x$d$\Theta_y$)] rad$^{-2}$  vs $\Theta_x$ and $\Theta_y$.  The color bar in (c) applies for (a) and (b) as well.
(d)-(f): 
d$\tilde{N}_e$/d$\Theta_x$  vs $\Theta_x$ corresponding to (a)-(c), respectively. The electron dynamics are simulated via ((a) and (d)) MCM, ((b) and (e)) MLLM, and ((c) and (f)) LLM, respectively. The collision angle of the electron beam $\theta_e=175^{\circ}$, and other laser and electron parameters are the same as in Fig.~\ref{fig2}.}
	\label{fig6}
\end{figure}

\section{The impacts of laser and electron parameters on the SE signature}

We have further investigated the impacts of the laser pulse and electron beam parameters on the SE signature in the electron angular distribution. 
For experimental feasibility, we first consider the case of a large energy spread of the electron beam.
The results in the case of a large  energy spread $\Delta\varepsilon_i/\varepsilon_i=0.1$  show a stable SE signature compared with those in Fig.~\ref{fig2}, see Fig.~\ref{fig4}.
And, we also investigate the cases with a collision angle $\theta_e=179^\circ$ in Fig.
~\ref{fig5} and $\theta_e=175^\circ$ in Fig.~\ref{fig6}.  Comparing  Fig.~\ref{fig5} with Fig.~\ref{fig2}, as $\theta_e$ shifts 1$^\circ$ from 180$^\circ$ to 179$^\circ$, in MCM the electron density peak moves left about $6^\circ$; in MLLM and LLM the electron density peaks in the left rise, and those in the right decline, respectively. However, the electron distribution signature  is similar to that in Fig.~\ref{fig2}.
Moreover, comparing  Fig.~\ref{fig6} with Fig.~\ref{fig2}, as $\theta_e$ shifts 5$^\circ$ from 180$^\circ$ to 175$^\circ$, the electrons deposit mainly in the  region of $\Theta_x < 0^\circ$. In MCM, the electrons deposit in a sub-elliptial region in the angle-resolved electron distribution with one peak close to the center, see Fig.~\ref{fig6}(a). In MLLM and LLM, the electrons both deposit in a fan-shaped region with one peak at the edge, see Figs.~\ref{fig6}(b) and \ref{fig6}(c). However, the distinctions between the three models are still obvious, see also Figs.~\ref{fig6}(d)-\ref{fig6}(f).
Thus, the expected radiative aggregation dynamics of electrons are clearly distinguishable.

  \begin{figure}[t]
 	\includegraphics[width=1.0\linewidth]{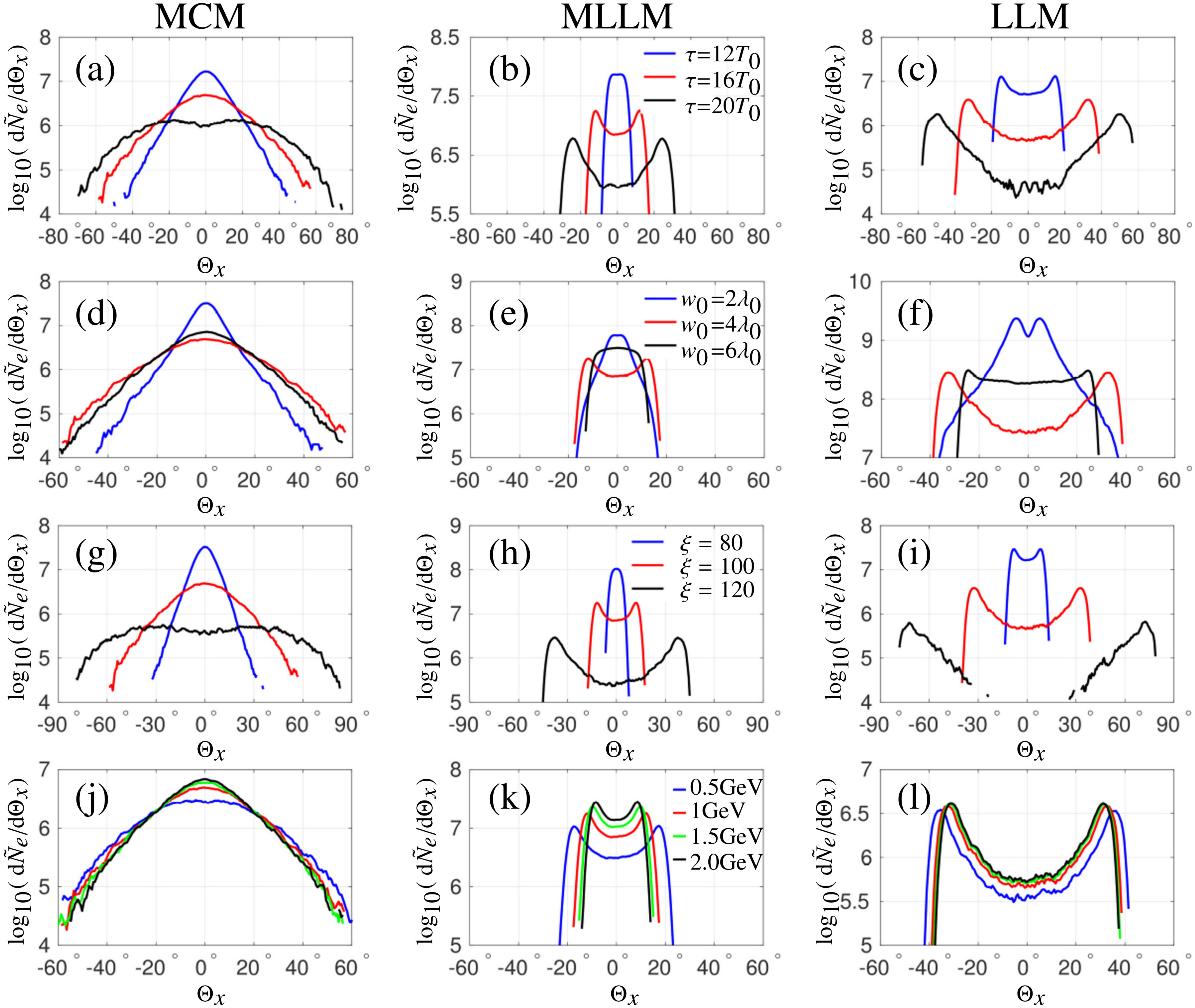}
 	\caption{Impacts of (a)-(c) the pulse duration, (d)-(f) the focal radius, and (g)-(i) the peak intensity of the laser pulse, as well as (j)-(l) the initial kinetic energy of the electron beam on the angle-resolved electron-number distributions. The simulation models are MCM (left column), MLLM (middle column), and LLM (right column), respectively. Other laser and electron parameters are the same as in Fig.~\ref{fig2}.  }
 	\label{fig7}
 \end{figure}

Furthermore, the role of the laser pulse duration $\tau$ is analyzed in Figs.~\ref{fig7}(a)-\ref{fig7}(c). As  $\tau$ increases from 12$T_0$ to 20$T_0$, the laser-electron interaction time increases gradually, which allows the ponderomotive force to deflect the electrons further outwards. Consequently, the peak strength via MCM declines, and the angular radii of the ``ring'' structures in MLLM and LLM both increase. For clear SE one should choose an intermediate laser pulse duration. In fact, as the laser pulse duration, i.e., the laser-electron interaction time, is too long, in MCM the stochastic-radiation aggregation effect of electrons could not overcome the electron-beam-expansion effect due to the ponderomotive force. On the contrary, if the laser pulse duration is too short, via MLLM and LLM the ponderomotive force cannot deflect the electrons  outwards enough to form the ``ring'' structure.

 The role of the laser focal radius is analysed in Figs.~\ref{fig7}(d)-\ref{fig7}(f). The latter show that the case $w_0\approx 2w_e$ is optimal for the observation of SE.
  When $w_0 =w_e=2\lambda_0$,
  electrons near the electron-beam boundaries experience rather weak laser fields, can not be deflected outwards much, and consequently, keep their initial motion directions near  $\Theta_x=0^\circ$ for all three models. However, when $w_0$ increases to $3w_e$, the laser-intensity gradient on the cross section of the electron beam becomes much smaller,  and the laser ponderomotive force $F_p\propto\bigtriangledown|E^2|$ is rather weak accordingly.  Thus, the deflection effects are weakened, and the electron angular distributions in Figs.~\ref{fig7}(e) and \ref{fig7}(f) vary little from the center to the peripheries.

The laser peak intensity can remarkably affect the electron dynamics, as shown in Figs.~\ref{fig7}(g)-\ref{fig7}(i). As $\xi$ increases from 80 to 120, the electron density near $\Theta_x=0^\circ$ decreases apparently, since the ponderomotive force increases. However, the distinctions among the three models are obvious.  Furthermore, the initial kinetic energy of the electron beam does not evidently affect the electron
 distribution, see Figs.~\ref{fig7}(j)-\ref{fig7}(l). As $\varepsilon_i$ increases from 0.5 GeV to 2 GeV,  $\theta_d $ decreases accordingly.
We find that the electron aggregation effect is more obvious when the condition of   $\xi/\gamma \sim 1/20$ is fulfilled.

Thus, the qualitative SE signature is easily observable at current achievable experimental conditions of the laser and electron beam.  

\section{Conclusion}

In conclusion, we have investigated  SE of photon emission on the dynamics of an electron beam colliding head-on with an ultraintense focused circularly-polarized laser pulse in the quantum radiation-dominated regime with the condition of $\gamma\sim 20 \xi$. Due to SE the electrons aggregate
inwards to the laser propagation axis, resulting in
a peak structure in electron angular distribution near the beam propagation direction, with a FWHM of tens of degrees.
This is in contrast to the case without SE, when
the ponderomotive force of the laser fields will deflect the electrons outwards and generates a
``ring'' structure in the angular distribution with a spread of about $10^\circ$. The
SE signature is very robust with regard to the laser pulse and the electron beam parameters in currently available laser facilities.

\section{Acknowledgments}
This work is supported by the  Science Challenge Project of China (No. TZ2016099), the National Key Research and Development Program of China (Grant No. 2018YFA0404801), and the National Natural Science Foundation of China (Grants Nos. 11874295, 11804269, U1532263). 

\bibliography{SE}

\end{document}